# 4π Planning for the Reduction of Predicted Hematologic Toxicity Risk in Cervical Cancer Radiotherapy

Haotian Feng[1,*], Yan Kong[1,2,*], Qifan Xu[1], Ke Sheng[1,**]

[1]Dept. of Radiation Oncology, University of California-San Francisco, San Francisco, United States
[2]Dept. of Radiation Oncology, Affiliated Hospital of Jiangnan University, Wuxi, Jiangsu, China
* These authors have equal contributions.
** Corresponding author: Ke.Sheng@ucsf.edu

## ABSTRACT

**Purpose:** In conventional coplanar radiotherapy for cervical cancer, nearby pelvic bones receive high radiation doses, increasing the risk of acute hematologic toxicity (HT). This study aims to estimate the HT risk reduction achievable with non-coplanar (4π) radiotherapy.

**Methods:** We retrospectively analyzed 114 cervical cancer patients treated with coplanar volumetric modulated arc therapy (VMAT) between 2021 and 2023. Radiomic features from planning CTs were extracted and combined with clinical and dosimetric data to train predictive machine learning models. Patients were then replanned using integrated non-coplanar beam orientation and fluence map optimization (4π planning). The top-performing model was applied to these new plans to evaluate potential HT reduction.

**Results:** Models combined radiomics, clinical, and dosimetric features achieved the highest predictive performance (AUC = 0.79). Feature importance analysis highlighted CT radiomic and dosimetric variables as the strongest predictors. Compared to VMAT, 4π non-coplanar planning significantly reduced doses to the bone marrow ($V_{10Gy}$, $V_{20Gy}$, $V_{30Gy}$, and $V_{40Gy}$ by 28%, 52%, 47%, and 33%, respectively) while significantly reducing dose to other pelvic organs at risk (OAR). When evaluated by the model, this improved dosimetry translated into a 23% reduction in predicted HT risk (risk ratio: 0.77; 95% CI: 0.75–0.79 and odds ratio: 0.68; 95% CI:0.65–0.71).

**Conclusions:** Non-coplanar 4π radiotherapy significantly lowers radiation doses to major pelvic bones during cervical cancer treatment without compromising target coverage or sparing of other OARs. Based on our predictive model, this superior dosimetry should translate to a marked reduction in acute hematologic toxicity.

# 1. Introduction

Cervical cancer remains one of the most prevalent malignancies affecting women worldwide, particularly in low and middle-income countries(1). External beam radiation therapy (EBRT), often delivered in combination with brachytherapy and chemotherapy, is a cornerstone in the management of locally advanced disease(2). However, EBRT inevitably exposes surrounding normal tissues to radiation, including the pelvic bone marrow (BM), which contains more than half of the body's active hematopoietic reserve(3).

The pelvic BM, distributed across the ilium, sacrum, proximal femur, and lower lumbar spine, is highly radiosensitive. Radiation-induced BM injury impairs hematopoietic stem cell function and disrupts the microenvironment necessary for hematopoiesis, leading to acute and chronic hematologic toxicity (HT)(4). These toxicities often manifest as leukopenia, anemia, and thrombocytopenia, which not only interrupt concurrent chemoradiotherapy(5) but also compromise therapeutic efficacy and patient outcomes(6). Interindividual variability in BM radiosensitivity and hematopoietic reserve further complicates toxicity prediction, underscoring the need for personalized risk assessment tools(7).

Coplanar volumetric modulated arc therapy (VMAT) is the current clinical standard for cervical cancer radiotherapy, offering efficient delivery with good target coverage(8, 9). However, OAR doses remain high due to challenging anatomy. To irradiate the target, the coplanar X-rays inevitably overlap their entrance and exit paths in the OARs. In comparison, non-coplanar 4π radiotherapy with optimized beam angles can explore a larger beam solution space, thereby minimizing such overlap. The efficacy of 4π planning has been demonstrated in the liver(10, 11), lung(12), prostate(13), brain(14), and head-neck cases(15–19). It is of significant clinical interest to estimate the potential reduction of HT using 4π non-coplanar planning. The recently developed ultra-high-performance parallel (UHPP) 4π system has overcome the computational challenges(20), enabling high-throughput noncoplanar optimization for the large PTV volume in cervical treatment. A subsequent question is whether the improved dosimetry can be translated into meaningful HT reduction.

Traditional dose-volume histogram (DVH) metrics are known to correlate with radiation-induced bone marrow (BM) toxicity(3, 21, 22). Specifically, a bone marrow volume receiving 40 Gy or more is significantly associated with HT(23). However, relying solely on the total bone marrow dose fails to account for the heterogeneous marrow reserve among patients. Within bone marrow, the active marrow—which can be identified using molecular imaging techniques like FLT and decreases with age and prior chemoradiotherapy(24)—has been the focus of several dosimetric studies(25, 26). These studies demonstrate that the radiation dose delivered specifically to the active bone marrow is a more accurate predictor of HT(27).

Despite its clinical value, implementing active bone marrow-sparing (BMS) via functional imaging faces significant bottlenecks: the molecular tracers required to detect proliferation lack FDA approval and are prohibitively expensive for widespread adoption. While magnetic resonance imaging (MRI) offers an alternative by inferring active marrow volumes from water-to-fat ratios, its quantitative accuracy is limited(28). Moreover, the dose threshold for functional sparing has not been clearly established(29).

As an alternative to the explicit approach, implicit data-driven methods have been adopted to incorporate the additional imaging features. Radiomic features extracted from planning CT may capture subtle variations in tissue microarchitecture associated with hematopoietic reserve. Subsequently, significantly improved HT prediction accuracy has been reported when radiomics is added to the model(30, 31). It remains a question how CT images reflect the individual hematopoietic reserve, which is insensitive to contrast between active and inactive bone marrow. One possible explanation is that bone mineral structures, which CT is sensitive to, show a clear inverse, site-specific correlation with active and inactive bone marrow, reflecting the same mesenchymal stem cell origin of osteoblasts and adipocytes(32).

In the current study, we first comprehensively evaluate HT prediction models using imaging, dosimetric, and clinical inputs, and then select the top-performing model to quantify the HT risk reduction achieved with 4π non-coplanar radiotherapy for cervical cancer treatment.

# 2. Methods

This study adhered to the TRIPOD statement(33) to ensure transparent reporting of model development, validation, and performance assessment (**Supplementary Material**). An overview of whole pipeline is shown in **Figure 1**.

## 2.1 Dataset

This retrospective study included 114 consecutive patients with cervical cancer who underwent pelvic external beam radiotherapy (EBRT) at Jiangnan University Affiliated Hospital between 2021 and 2023. To isolate the effect of radiotherapy on HT, patients who received concurrent or adjuvant chemotherapy were excluded. Acute HT was defined as grade ≥2 toxicity according to the Common Terminology Criteria for Adverse Events (CTCAE v5.0) based on complete blood count measurements obtained before, during, and within one month after radiotherapy. Among the 114 patients, 54 (47%) developed acute grade ≥2 HT. Detailed patient eligibility criteria, treatment delivery, and toxicity assessment are provided in the **Appendix A**.

## 2.2 Pelvic Bone Delineation, Feature Extraction and Selection

Pelvic regions of interests (ROIs) were delineated on treatment planning CT images using 3D Slicer software. In this study, the pelvic bony contours containing BM were delineated into three subregions. Later, radiomic (through PyRadiomics(34)), dosimetric, and clinical features were extracted and used to train the prediction model regarding HT. Additional delineation and implementation details are provided in the **Appendix B**.

## 2.3 Treatment Optimization for 4π non-coplanar

To improve the PTV-OAR dose gradient, 4π non-coplanar planning was optimized using FISTA and then normalized to ensure 95% PTV coverage, consistent with clinical VMAT plans. The dose was calculated using collapsed-cone convolution with Monte Carlo validation(20). The PTV dose was set to 50 Gy, and the OAR limits matched those of the VMAT plan (see **Appendix C**).

Optimization began with 1,162 beams spaced 6° apart, which was reduced to ~400 candidates after eliminating collisions and CT boundary intersections. System parameters included a 100 cm SID, a 0.5 cm beamlet size, and a 20 cm × 20 cm field size due to a GPU memory limitation.

### 2.4 Multi-isocenter PTV Region Split

To accommodate a maximum field size of 20 cm, the cervical Planning Target Volume (PTV) needed to be partitioned across multiple isocenters. This was achieved by computing a 3D bounding box around the PTV and bisecting its two longest axes to create four equal-volume sub-cuboids (**Figure 2**). The shortest axis remained undivided, as it naturally fits within the parameters of a single field. Within each sub-cuboid, the center of mass of the enclosed PTV voxels was calculated. Because of the PTV's irregular morphology, these tissue centroids deviate from the sub-cuboids' geometric centers; using the true tissue centroids for isocenter placement ensures precise targeting based on the actual volume distribution. Finally, **Figure 2c** illustrates the resulting four contiguous, non-overlapping sub-regions mapped directly onto the tumor surface.

## 3. Results

### 3.1. Univariate Analysis of Demographic and Dosimetric Features

We first evaluated the statistical significance of the clinical features based on their p-values. The Chi-squared ($\chi^2$) test was applied to categorical variables (FIGO stage and pathology type), while continuous variables (age and BMI) were analyzed using either the independent t-test or the Mann-Whitney U test, depending on normality, as assessed by the Shapiro-Wilk test(35). The summary statistics and corresponding p-values of demographic and dosimetric features were presented in **Appendix D** and **E**.

A univariate analysis of dosimetric features with respect to toxicity risk was performed for each anatomical structure, with results shown in **Figure 3(a-b)**. Across three regions and whole pelvic bones, $D_{x\%}$ and mean dose show consistently positive coefficients, indicating that higher doses are associated with increased toxicity risk — a finding consistent with general dose-response expectations. Similarly, most key dose-volume features ($V_{20Gy}$ through $V_{45Gy}$, $D_{1\%}$, $D_{2\%}$) exhibit positive coefficients across all four regions, suggesting that increased volume exposed to intermediate-to-high dose thresholds is broadly associated with elevated risk.

Notably, $V_{50Gy}$ showed a negative coefficient in the lower pelvis, lumbosacral vertebrae, and iliac bone (panels a, b, d), and $V_{5Gy}$ and $V_{10Gy}$ were likewise negative in select structures (panels a and d). This sign inversion at the low- and high-dose extremes was inconsistent with the otherwise monotonic dose-response pattern observed across the mid-range dose-volume features and likely reflected collinearity among dose-volume metrics rather than a genuine protective effect — a limitation we addressed further in the multivariate analysis. We also noted several coefficients with large magnitudes relative to the rest of the panel (e.g., $V_{5Gy}$ = -53 in the iliac bone), which might indicate feature instability driven by limited sample size or a narrow dynamic range in those specific dose-volume bins for the corresponding structure.

### 3.3 Hematologic Toxicity Prediction

To evaluate predictive effectiveness, we analyzed combinations of radiomic, demographic, and dosimetric features. **Figure 3(c-f)** details the resulting AUC values across various machine learning algorithms. Clinical and dosimetric features used in conventional predictive models yielded the lowest performance (AUC<0.60). The addition of radiomics features resulted in the highest predictive performance and feature efficiency. Both elastic-net logistic regression (**Figure 3c**) and a neural network (**Figure 3f**) achieved a peak AUC of 0.79. While the neural network exhibited the best sensitivity-specificity balance for capturing complex relationships, we ultimately selected the logistic regression model to prioritize feature interpretability and model robustness. Notably, the additional contribution of dosiomic features from dose images was negligible, consistent with a previous study(36). Therefore, dosiomic features were excluded from the prediction model.

## 3.4 4π Dose Plan Optimization

**Figure 4** illustrates an example of a cervical cancer patient treated using the 4π plan. The cervical region was divided into four subdomains, and four isocenters were employed to optimize the treatment plan. Multi-isocenter optimization has been shown to be dosimetrically desirable for treating large targets with high modulation resolution(37). **Figure 4a** showed a cutout view centered on the cervical region. From these images, it is evident that the dose delivered to OARs was substantially lower with the 4π plan than with the original VMAT plan, while maintaining adequate target coverage.

**Figure 4b** compares the DVH profiles of a representative patient's 4π and VMAT plans. 4π maintained comparable PTV coverage while notably reducing radiation to surrounding OARs. While 4π produced a substantial dose reduction in targeting bone regions and OARs, including the bladder, femoral head, and bowel bag. **Figure 4c** further presented the dose distributions across different bone regions for VMAT and 4π plans. The VMAT plan demonstrated relatively extensive regions of moderate- and high-dose exposure in the lower pelvis, bilateral iliac bones, and lumbosacral region. In contrast, the 4π plan substantially reduced both the spatial extent and the intensity of dose deposition, with most of the bone volume receiving low doses. The most pronounced visual difference was observed in the lumbosacral region, where the broad moderate-dose distribution seen with VMAT was markedly reduced with 4π planning. Overall, these representative dose maps visually demonstrated the potential of 4π optimization to reduce radiation exposure to pelvic bone marrow while maintaining the prescribed treatment objectives.

**Figure 5** summarizes key dosimetric feature distributions across all patients for bone regions and OARs, respectively, with full statistical comparisons detailed in **Appendix F**. Overall, switching from VMAT to 4π reduced most dosimetric parameters in the target regions. Notably, normalized total pelvic bone $V_{10Gy}$, $V_{20Gy}$, $V_{30Gy}$ and $V_{40Gy}$ reduced from 93%, 75%, 45%, and 24% to 67%, 36%, 24%, and 16%, respectively (all $p < 0.05$). Significant dosimetric improvements were also observed for targeting OARs. For example, rectum $V_{30Gy}$ was reduced by 5%, bladder $V_{45Gy}$ by 2%, small intestine $V_{40Gy}$ by 2%, and bowel bag $V_{35Gy}$ by 2% (all $p <$ 0.05). Collectively, these findings confirmed that 4π non-coplanar robustly and consistently achieved significant dose reductions compared to conventional coplanar VMAT planning for cervical RT.

### 3.5 Risk Comparison between VMAT and 4π plan

To evaluate the 4π plan, we applied our trained radiomic model to predict HT probabilities for both the 4π and VMAT plans, comparing them via a Risk Ratio (RR). We assessed uncertainty using two bootstrapping methods. **Full-pipeline bootstrapping** (accounting for both training and prediction uncertainty) yielded an RR of 0.80 (95% CI: 0.57–1.03); because this broad interval crosses 1.0, the risk reduction is not statistically significant when combined. Conversely, **fixed-model bootstrapping** isolated prediction uncertainty, producing a narrower RR of 0.88 (95% CI: 0.87–0.90). Finally, a direct point estimate without bootstrapping (**Figure 6**) yielded an RR of 0.77 (95% CI: 0.75–0.79). Consistent with this finding, the corresponding odds ratio was 0.68 (95% CI: 0.65–0.71). Ultimately, all estimates consistently indicate a reduced HT risk with 4π-based optimization.

## 4. Discussions

In this study, we created fully integrated, non-coplanar plans for cervical cancer external beam radiation therapy and compared them with standard-of-care coplanar VMAT plans. We demonstrated markedly reduced dose to the pelvic bones and small bowel. Beyond these dosimetric improvements, we quantified the reduction in HT using a prediction model integrating imaging, dosimetric, and clinical features. This study differs from previous cervical radiotherapy research, which focused on dosimetric endpoints. Prior clinical studies—such as those by Mell et al.(38), Rose et al.(39), and the systematic review by Corbeau et al.(40)—have established that reducing specific dose volume metrics (e.g., $V_{10Gy}$, $V_{20Gy}$) in the pelvic bone minimizes Grade 2+ leukopenia and neutropenia. However, the dosimetric metrics derived at the population and statistical levels do not reflect individual patients' resilience to radiation, which is further influenced by the underlying hematopoietic reserve, the distribution of active bone marrow, and intrinsic radiosensitivity. Therefore, optimizing dosimetric parameters alone may not translate into personalized bone marrow sparing or HT risk reduction. Recently, imaging features extracted from readily available planning CTs have been incorporated into HT prediction, resulting in markedly improved accuracies(30, 31). By integrating these two research directions, our approach directly estimates the most relevant clinical endpoints and uses advanced planning techniques to provide active decision support.

Our risk prediction model demonstrated that integrating CT radiomic features with clinical and dosimetric data substantially outperforms models that rely solely on dose and clinical factors. The predictive performance of our unified model aligns with that of existing radiomics-based HT models for cervical cancer(30, 31), with one distinction: the previous studies included chemoradiotherapy patients, whereas our cohort consisted exclusively of RT patients.

Due to the importance of HT in cervical cancer patient management, bone marrow sparing planning has been extensively researched, which is summarized in a meta-analysis(41). Coplanar VMAT inevitably causes overlap between the entrance and exit beams, which is particularly detrimental to the pelvic PTV sandwiched between the bone marrow and the small intestine(42). Aggressive marrow sparing increases dose spillage into the small intestine and compromises target coverage. As a result, only moderate reductions (<10-15% in most cases) in the bone marrow dose $V_{10Gy}$ to $V_{40Gy}$ were achieved in 22 clinical studies(41) when additional dose constraints were applied to the pelvic bones in coplanar planning, compared with 28%-52% achieved using 4π non-coplanar planning in this study. The 4π framework achieves a significantly greater dose reduction by exploring a much larger solution space(20). Using a conical beam pattern, the non-coplanar approach minimizes overlap between the entrance and exit beams. Our trained model further predicts a 23% reduction in the risk of acute HT.

In practice, this predictive pipeline supports a targeted, resource-conscious clinical triage strategy. Individuals flagged by the model as having a high HT risk under standard VMAT can be prioritized for advanced 4π planning. Conversely, patients with a low predicted risk can continue with standard VMAT to benefit from its faster optimization and delivery times. This selective approach ensures that the increased practical demands of non-coplanar 4π delivery—such as longer optimization times, multiple couch rotations for C-arm systems, and rigorous collision-avoidance quality assurance(43)—are concentrated only on the patients most likely to benefit. Because the risk-prediction model relies entirely on standard, pre-existing CT and dosimetric datasets, the predictive workflow can be seamlessly integrated into current clinical routines without requiring specialized imaging or invasive testing, even if the eventual 4π delivery demands greater clinical resources.

Despite the encouraging results, several limitations merit discussion. First, the current study is based on a single-institution retrospective cohort. Future work will involve multi-center validation and incorporation of external datasets to assess robustness across scanners, reconstruction parameters, and patient populations. Second, even with univariate feature selection, our relatively small sample size compared to the vast pool of candidate radiomic features introduces a risk of residual overfitting. This risk is particularly relevant to the benchmarked neural network, which motivates our primary reliance on the more parsimonious elastic-net model. Larger future cohorts are needed to determine whether the neural network's added complexity yields genuine predictive gains or merely fits cohort-specific noise. Third, our estimated risk reduction is highly sensitive to how uncertainty is propagated through the bootstrap procedure. When isolating prediction-stage uncertainty, the estimated risk ratio is 0.77 (95% CI: 0.75–0.79). However, when accounting for full model-refitting uncertainty, the interval widens to 0.80 (95% CI: 0.57–1.03). We report both intervals for transparency, while future studies with larger sample sizes should provide more robust estimation. Fourth, because the model was trained on historical VMAT plans and then applied to 4π dose distributions, it had to extrapolate beyond the dosimetric range it encountered during training. As a result, the model likely underestimated the clinical benefits of the substantially lower marrow doses achieved with 4π. This underestimation is apparent when comparing our results to a meta-analysis of 11 randomized BMS radiotherapy trials(41): although 4π reduces pelvic bone doses significantly more than the techniques in those trials, our model predicted a comparable, rather than proportionally greater, risk reduction for G2+ HT (odds ratio=0.27). Another potential contributor to the apparent discrepancy is chemotherapy, which is absent in the current study but could amplify the HT in the clinical trial cohorts. Finally, although the workflow relies only on standard planning CTs, evidence linking CT to bone marrow distribution remains indirect(32). Future research should further interrogate the correlation between CT and functional images.

# 5. Conclusion

This study proposes a closed-loop precision radiotherapy framework that uses a toxicity model to quantitatively evaluate and personalize advanced treatment planning. We introduced an optimization strategy based on the 4π non-coplanar framework to improve the dose gradient and subsequently the sparing of pelvic bone marrow in cervical external beam radiotherapy. The potential efficacy of reducing HT is further demonstrated using an individualized predictive model that combines imaging and clinical features.

## Author Contributions

Conceptualization, K.S.; methodology, H.F., K.Y., Q.X.; investigation, H.F., K.Y., Q.X. and K.S.; writing-original draft, H.F., K.Y., Q.X. and K.S.; funding acquisition, K.S.; supervision, K.S.

## Declaration of Interests

The authors declare that they have no known competing financial interests or personal relationships that could have appeared to influence the work reported in this paper.

## Data Availability

The data is available upon reasonable request to the corresponding author.

## Funding Statement

This work is supported by institutional funding.

# Figure Legends

**Figure 1**. Machine Learning workflow and treatment planning pipeline for cervical cancer toxicity risk estimation.

**Figure 2. (a)** Example of one patient under the 4π dose plan. **(b)** The whole PTV (light green surface) is enclosed by its bounding box and divided into four sub-cuboids (red, blue, green, magenta outlines) by bisecting the box along its two longest axes. Colored markers indicate the centroid of the PTV sub-volume within each sub-cuboid, used as the isocenter for that region. **(c)** The same four-way split is visualized directly on the PTV surface, with each color denoting the portion of the target assigned to the corresponding isocenter.

**Figure 3.** (a-b) Univariate analysis of dosimetric parameters with respect to toxicity label, shown separately for four anatomical structures. Positive coefficients indicate increasing risk with increasing dose/volume; negative coefficients indicate the opposite trend. Numeric annotations denote coefficients that exceed the plotted axis range. (c-f) HT prediction accuracy between different machine learning models: (c) logistic regression with elastic-net; (d) random forest; (e) gradient boosting; (f) neural network. 'CT' denotes radiomic features extracted from the CT image, 'Clinical' represents the demographic and clinical features, and 'Dosimetric' denotes dosimetric features from DVH.

**Figure 4.** (a) 2D centroid cutout view of one patient between VMAT plan and 4π plan in cervical cancer radiation therapy. The contours represent the locations of different OARs. (b) Comparison of DVH profiles for VMAT (dash line) and 4π (solid line) for one patient. (c) Dose distribution at different bone regions.

**Figure 5**. Boxplot comparison between VMAT and 4π over different regions (lower pelvis, iliac bone, lumbosacral vertebrae, and all regions (whole pelvic) regarding: (a) $V_{30Gy}$; (b) $V_{40Gy}$; (c) $D_{1\%}$.

**Figure 6**. Risk probability and Risk Ratio (RR) between VMAT and 4π dose plans for all patient cohort: (a) risk probability between different plans; (b) risk ratio between different plans where baseline VMAT risk ratio is set to 1.

# Appendix

## A. Detailed Dataset Information

### Patients

This retrospective study included 114 patients with cervical cancer who underwent pelvic EBRT at Jiangnan University Affiliated Hospital between 2021 and 2023. To specifically investigate the impact of radiotherapy alone on HT, patients who received adjuvant chemotherapy were excluded. The inclusion criteria were as follows: pathologically confirmed cervical cancer; absence of distant metastasis; no prior history of pelvic radiotherapy; no concurrent chemotherapy; complete blood count (CBC) performed within one week before radiotherapy, during treatment, and one month after treatment; absence of ≥ grade 1 HT or long-term anemia before radiotherapy. Patients with other malignancies, recurrent tumors, or hematologic disorders were excluded.

### Radiotherapy delivery

All patients underwent CT simulation and subsequent treatment in the supine position with a full bladder, achieved by drinking 1 liter of water 1 hour before simulation. CT scans were performed using a LightSpeed Ultra scanner (GE Healthcare, USA) with the following parameters: slice thickness 5 mm, tube voltage 120 kVp, automatic tube current, and matrix size 512 × 512. The scanning range extended from the first lumbar vertebra to the mid-shaft of the femur. EBRT was delivered using volumetric modulated arc therapy (VMAT), consisting of two coplanar mirrored arcs. Treatment planning was performed in the Varian Eclipse system (version 13.6) with the Acuros XB (AXB) algorithm, and the plans were delivered on Varian VitalBeam. The prescribed dose to the planning target volume (PTV) was 45-50.4 Gy in 25-28 fractions. BM was not delineated as an avoidance structure in any treatment plan.

### Hematologic toxicity evaluation

Complete blood count data, including white blood cell (WBC) count, absolute neutrophil count (ANC), hemoglobin (HGB), and platelet (PLT) levels, were collected from 1 week before the initiation of radiotherapy to 1 month after its completion(34). HT was graded according to the Common Terminology Criteria for Adverse Events (CTCAE), version 5.0(35). Acute HTs were defined as those occurring within 1 month of completing radiotherapy. A toxicity of Grade ≥ 2 (G2+) in WBC, ANC, HGB, or PLT was defined as an acute HT event and was considered a positive endpoint in the subsequent analysis. Among the 114 patients included in this study, among which 54 (47%) developed acute G2+ HT during radiotherapy.

## B. Pelvic Bone Delineation and Feature Extraction

The pelvis was defined as extending from the upper edge of the L5 vertebral body to the inferior border of the ischial tuberosities and was further divided into three subregions: (1) Iliac bone, extending from the iliac crest to the superior edge of the femoral head; (2) Lower pelvis, extending from the superior edge of the femoral head to the inferior edge of the ischial tuberosity; and (3) Lumbosacral vertebrae, extending from the upper edge of L5 to the entire sacrum. Delineation was performed independently by two radiation oncologists to ensure accuracy. An example of the relative locations is shown in **Figure B1**.

We collected three categories of features: radiomic, demographic, and dosimetric. To capture tumor heterogeneity, we used PyRadiomics to extract original and wavelet-transformed shape, first-order, and texture features (GLCM, GLDM, GLRLM, GLSZM, and NGTDM) from CT images. Demographic data included age, BMI, pathological type, FIGO stage, and complete blood counts (WBC, ANC, HGB, and PLT). Finally, we extracted dosimetric features from the DVH—specifically mean dose, $V_{xGy}$, and $D_{x\%}$. These parameters were calculated across three regions: the lumbosacral vertebrae, lower pelvis and ilium.

Radiomic and statistical features from the three bone regions were combined for each patient. To identify the most predictive features for HT, we applied the SelectKBest algorithm using an ANOVA F-test. We then trained an elastic-net logistic regression model, optimizing regularization parameters based on prediction performance. Evaluation utilized a 5-fold cross-validation strategy (60%/20%/20% train-validation-test for each fold), with preprocessing strictly confined to the training folds to prevent data leakage. Finally, predictive performance was benchmarked against three baselines: a 100-estimator random forest, gradient boosting, and a 2-layer neural network (32 neurons per hidden layer). The trained elastic-net logistic regression model was then applied to each patient's non-coplanar 4π (UHPP) plan and coplanar VMAT plan, generating a predicted HT probability for each plan type. These paired predictions were used to compute based on risk ratio and odds ratio (4π vs. VMAT) for each patient, allowing direct comparison of predicted toxicity between the two planning approaches.

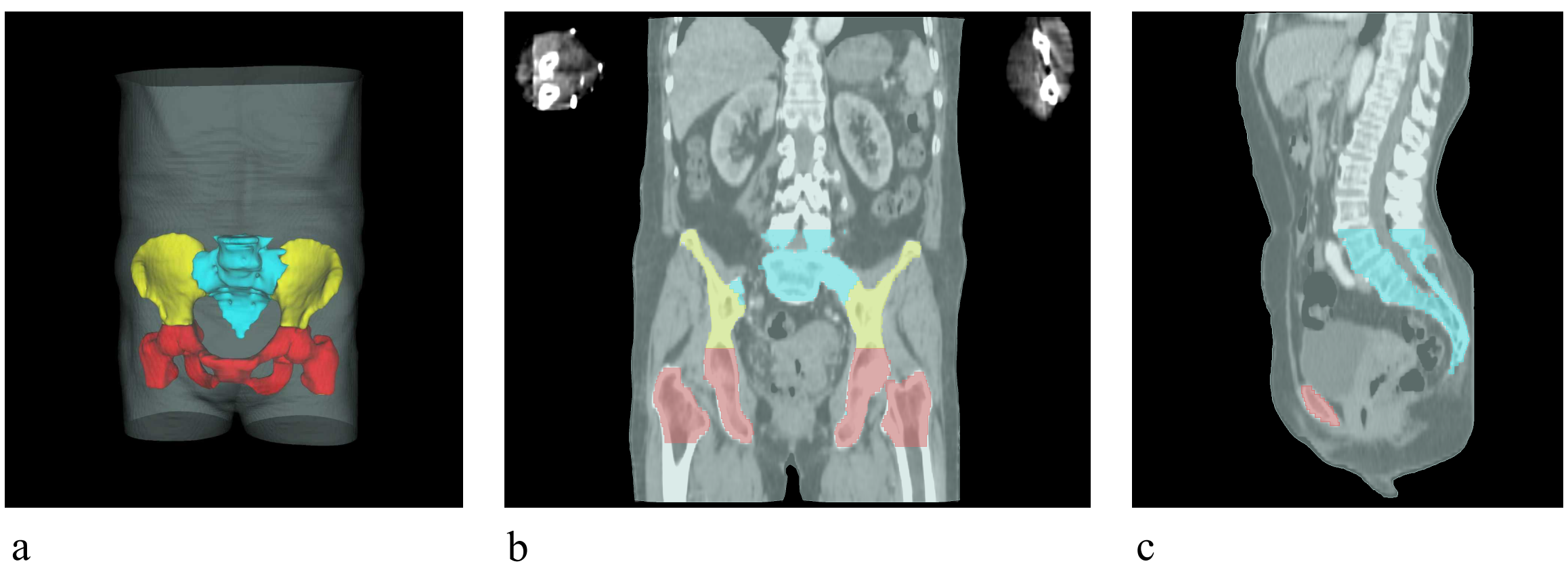

**Figure B1**. Spatial distribution of pelvic bone regions. Lower pelvis (red), iliac bone (yellow), and lumbosacral vertebrae (blue) are shown relative to the body. (a) 3D rendering of the body and segmented bone marrow masks. (b) Coronal view of CT image with overlaid masks. (c) Sagittal view of CT image with overlaid masks.

Given the large number of radiomic features extracted, we evaluated the performance of different feature-set sizes in predicting HT. **Figure B2** illustrated the relationships between the number of selected radiomic features based on the training set and the predictive performance across various machine learning algorithms. Our analysis showed that using 100 features provided the optimal balance between predictive accuracy and model complexity, achieving the best overall performance among the evaluated feature sets.

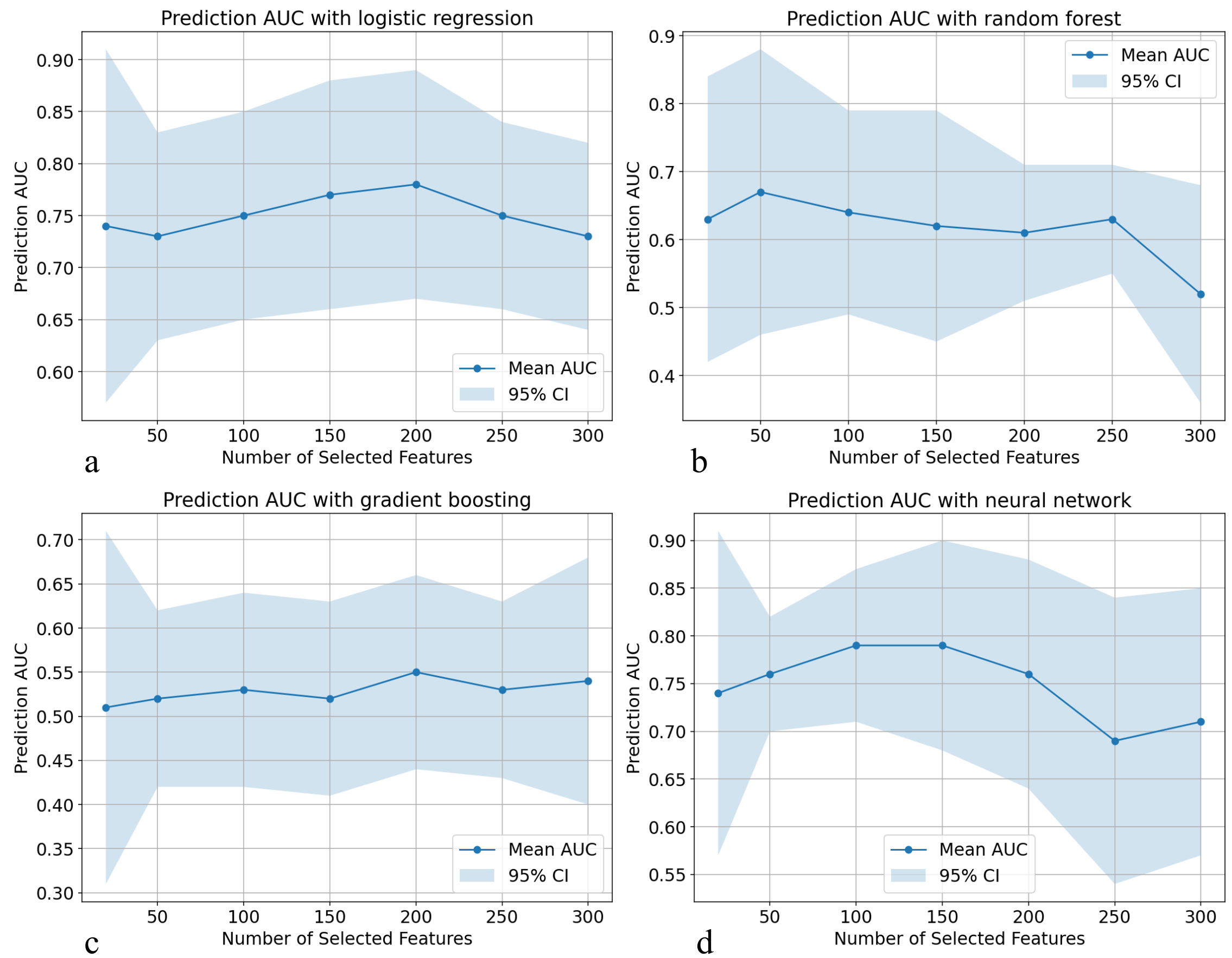

**Figure B2**. Prediction AUC with respect to different number of features (band denotes 95% confidence interval): (a) prediction AUC with logistic regression with elastic-net; (b) prediction AUC with random forest; (c) prediction AUC with gradient boosting; (d) prediction AUC with neural network.

## C. 4π plan

**Table C1** shows the detailed 4π plan parameters for PTV and OARs. Here, PTV + XMM/XCM denotes the region extending Xmm/Xcm outward from the planning target volume (PTV) to the body surface, defining the external dose-limiting boundary for beam selection and optimization.

**Table C1.** 4π plan optimization at different OARs

| NAME | MAX DOSE WEIGHTS | MAX DOSE (GY) | MIN DOSE TARGET WEIGHTS | MIN DOSE TARGET (GY) | OAR WEIGHTS | IDEAL DOSE (GY) |
|---|---|---|---|---|---|---|
| PTV 1 | 100 | 50 | 100 | 50 | -- | 50 |
| PTV 2 | 100 | 50 | 100 | 50 | -- | 50 |
| PTV 3 | 100 | 50 | 100 | 50 | -- | 50 |
| PTV 4 | 100 | 50 | 100 | 50 | -- | 50 |
| PTV + 1MM – 7MM RING | 10 | 40 | -- | -- | 10 | 0 |

| PTV + 2CM | 10 | 40 | -- | -- | 10 | 0 |
|---|---|---|---|---|---|---|
| PTV + 5CM | 10 | 40 | -- | -- | 10 | 0 |
| RECTUM | 10 | 40 | -- | -- | 10 | 0 |
| BLADDER | 10 | 40 | -- | -- | 10 | 0 |
| SMALL INTESTINE | 10 | 40 | -- | -- | 10 | 0 |
| BOWEL BAG | 10 | 40 | -- | -- | 10 | 0 |
| FEMUR HEAD | 10 | 40 | -- | -- | 10 | 0 |

## D. Univariate analysis of clinical features

**Table D1.** Clinical features and hematologic toxicity analysis

| | | HT NEGATIVE | HT POSITIVE | P-VALUE |
|---|---|---|---|---|
| AGE | | 61.72 ± 10.11 | 58.04 ± 12.31 | 0.08 |
| BMI | | 23.92 ± 3.93 | 23.20 ± 3.56 | 0.36 |
| WHITE BLOOD CELL ($\times 10^9$/L) | | 5.71 ± 1.53 | 5.77 ± 1.80 | 0.86 |
| NEUROPHILS ($\times 10^9$/L) | | 3.74 ± 1.37 | 3.71 ± 1.41 | 0.90 |
| RED BLOOD CELL ($\times 10^9$/L) | | 4.10 ± 0.42 | 4.17 ± 0.43 | 0.42 |
| HEMOGLOBIN (G/L) | | 122.29 ± 9.10 | 124.46 ± 14.11 | 0.37 |
| PLATELET ($\times 10^9$/L) | | 217.29 ± 58.91 | 225.96 ± 74.64 | 0.52 |
| FIGO STAGE | 1 | 15 | 8 | 0.35 |
| | 2 | 22 | 22 | |
| | 3 | 22 | 18 | |
| | 4 | 2 | 5 | |
| PATHOLOGY TYPE | 1 | 55 | 48 | 0.51 |
| | 2 | 3 | 6 | |
| | 3 | 1 | 1 | |

## E. Univariate analysis of dosimetric features

**Table E1.** Dosimetric features and hematological toxicity analysis

| REGION OF INTERESTS | DOSIMETRY | HT NEGATIVE | HT POSITIVE | P-VALUE |
|---|---|---|---|---|
| LOWER PELVIS | $V_{5Gy}$ | 0.92 ± 0.06 | 0.94 ± 0.05 | 5.33e-01 |
| | $V_{10Gy}$ | 0.81 ± 0.09 | 0.82 ± 0.09 | 6.00e-01 |
| | $V_{20Gy}$ | 0.54 ± 0.09 | 0.57 ± 0.08 | 1.84e-01 |
| | $V_{30Gy}$ | 0.30 ± 0.10 | 0.34 ± 0.12 | 1.51e-01 |
| | $V_{35Gy}$ | 0.21 ± 0.10 | 0.25 ± 0.12 | 2.41e-01 |
| | $V_{40Gy}$ | 0.14 ± 0.07 | 0.17 ± 0.09 | 1.78e-01 |
| | $V_{45Gy}$ | 0.07 ± 0.04 | 0.09 ± 0.04 | 5.16e-02 |
| | $V_{50Gy}$ | 0.01 ± 0.02 | 0.01 ± 0.01 | 4.46e-01 |
| | $D_{1\%}$ | 3.92 ± 0.04 | 3.92 ± 0.03 | 4.01e-01 |
| | $D_{2\%}$ | 3.90 ± 0.04 | 3.91 ± 0.03 | 3.74e-01 |
| | Mean Dose | 23.32 ± 3.53 | 24.25 ± 3.20 | 1.35e-01 |
| ILIAC BONE | $V_{5Gy}$ | 1.00 ± 0.00 | 1.00 ± 0.01 | 1.33e-01 |
| | $V_{10Gy}$ | 0.98 ± 0.02 | 0.98 ± 0.03 | 4.52e-01 |
| | $V_{20Gy}$ | 0.77 ± 0.10 | 0.79 ± 0.11 | 4.23e-01 |
| | $V_{30Gy}$ | 0.39 ± 0.12 | 0.40 ± 0.14 | 7.20e-01 |
| | $V_{35Gy}$ | 0.25 ± 0.09 | 0.27 ± 0.11 | 4.89e-01 |
| | $V_{40Gy}$ | 0.16 ± 0.06 | 0.17 ± 0.08 | 9.88e-01 |
| | $V_{45Gy}$ | 0.08 ± 0.04 | 0.09 ± 0.04 | 8.95e-01 |
| | $V_{50Gy}$ | 0.01 ± 0.02 | 0.01 ± 0.01 | 3.71e-01 |
| | $D_{1\%}$ | 3.92 ± 0.04 | 3.93 ± 0.03 | 3.70e-01 |
| | $D_{2\%}$ | 3.90 ± 0.04 | 3.91 ± 0.03 | 6.14e-01 |
| | Mean Dose | 28.14 ± 2.79 | 28.33 ± 2.81 | 7.21e-01 |
| LUMBOSACRAL VERTEBRAE | $V_{5Gy}$ | 1.00 ± 0.00 | 1.00 ± 0.00 | 5.84e-01 |
| | $V_{10Gy}$ | 1.00 ± 0.01 | 1.00 ± 0.00 | 2.81e-02 |
| | $V_{20Gy}$ | 0.95 ± 0.05 | 0.96 ± 0.03 | 1.62e-01 |
| | $V_{30Gy}$ | 0.70 ± 0.13 | 0.71 ± 0.12 | 6.85e-01 |
| | $V_{35Gy}$ | 0.54 ± 0.14 | 0.56 ± 0.13 | 6.34e-01 |
| | $V_{40Gy}$ | 0.40 ± 0.13 | 0.40 ± 0.11 | 7.51e-01 |
| | $V_{45Gy}$ | 0.24 ± 0.11 | 0.24 ± 0.09 | 1.00e+00 |
| | $V_{50Gy}$ | 0.05 ± 0.06 | 0.05 ± 0.06 | 5.49e-01 |
| | $D_{1\%}$ | 3.93 ± 0.04 | 3.94 ± 0.03 | 2.64e-01 |
| | $D_{2\%}$ | 3.92 ± 0.04 | 3.93 ± 0.03 | 3.17e-01 |
| | Mean Dose | 36.42 ± 3.18 | 36.98 ± 2.79 | 3.03e-01 |
| ALL REGION | $V_{5Gy}$ | 0.97 ± 0.02 | 0.98 ± 0.02 | 5.93e-01 |
| | $V_{10Gy}$ | 0.92 ± 0.04 | 0.93 ± 0.03 | 4.15e-01 |
| | $V_{20Gy}$ | 0.73 ± 0.05 | 0.75 ± 0.06 | 8.89e-02 |
| | $V_{30Gy}$ | 0.44 ± 0.09 | 0.46 ± 0.11 | 3.23e-01 |
| | $V_{35Gy}$ | 0.32 ± 0.08 | 0.34 ± 0.11 | 2.73e-01 |
| | $V_{40Gy}$ | 0.22 ± 0.06 | 0.23 ± 0.08 | 4.89e-01 |
| | $V_{45Gy}$ | 0.12 ± 0.04 | 0.13 ± 0.04 | 3.89e-01 |
| | $V_{50Gy}$ | 0.02 ± 0.03 | 0.02 ± 0.02 | 5.11e-01 |
| | $D_{1\%}$ | 3.93 ± 0.04 | 3.93 ± 0.03 | 3.11e-01 |
| | $D_{2\%}$ | 3.92 ± 0.04 | 3.92 ± 0.03 | 3.84e-01 |
| | Mean Dose | 28.82 ± 3.26 | 29.12 ± 2.34 | 5.69e-01 |

## F. Dosimetric feature comparison between VMAT and 4π

Dosimetric parameters obtained from the VMAT and 4π plans were compared using paired statistical tests. For each dosimetric metric, the normality of the paired differences was assessed using the Shapiro–Wilk test. If the paired differences were approximately normally distributed (Shapiro–Wilk $p > 0.05$), comparisons were performed using a paired *t*-test. Otherwise, the nonparametric Wilcoxon signed-rank test was applied. All statistical tests were two-sided, and a *p*-value $< 0.05$ was considered statistically significant.

**Table F1**. Dosimetric features at different bone regions between VMAT and 4π (P-Val represents P value)

| | Lower Pelvis | | | Iliac Bone | | | Lumbosacral Vertebrae | | | All Region | | |
|---|---|---|---|---|---|---|---|---|---|---|---|---|
| | VMAT | 4π | P-Val | VMAT | 4π | P-Val | VMAT | 4π | P-Val | VMAT | 4π | P-Val |
| $V_{5Gy}$ | 0.94 ± 0.05 | 0.81 ± 0.15 | 5e-17 | 1.00 ± 0.00 | 0.88 ± 0.16 | 2e-21 | 1.00 ± 0.00 | 0.93 ± 0.13 | 2e-20 | 0.97 ± 0.02 | 0.88 ± 0.11 | 6e-21 |
| $V_{10Gy}$ | 0.82 ± 0.09 | 0.54 ± 0.12 | 2e-21 | 0.98 ± 0.02 | 0.67 ± 0.18 | 2e-21 | 1.00 ± 0.01 | 0.82 ± 0.14 | 2e-21 | 0.93 ± 0.04 | 0.67 ± 0.12 | 2e-21 |
| $V_{20Gy}$ | 0.57 ± 0.11 | 0.26 ± 0.10 | 6e-57 | 0.78 ± 0.10 | 0.27 ± 0.12 | 1e-80 | 0.96 ± 0.04 | 0.60 ± 0.16 | 2e-21 | 0.75 ± 0.07 | 0.36 ± 0.11 | 2e-77 |
| $V_{30Gy}$ | 0.31 ± 0.11 | 0.17 ± 0.08 | 2e-21 | 0.38 ± 0.12 | 0.15 ± 0.08 | 2e-21 | 0.72 ± 0.13 | 0.45 ± 0.15 | 1e-43 | 0.45 ± 0.11 | 0.24 ± 0.09 | 4e-52 |
| $V_{35Gy}$ | 0.23 ± 0.11 | 0.14 ± 0.07 | 6e-20 | 0.25 ± 0.10 | 0.12 ± 0.06 | 4e-21 | 0.57 ± 0.14 | 0.38 ± 0.14 | 2e-19 | 0.33 ± 0.11 | 0.20 ± 0.08 | 1e-36 |
| $V_{40Gy}$ | 0.15 ± 0.08 | 0.11 ± 0.06 | 1e-14 | 0.16 ± 0.07 | 0.09 ± 0.05 | 4e-19 | 0.43 ± 0.13 | 0.31 ± 0.13 | 7e-16 | 0.24 ± 0.09 | 0.16 ± 0.07 | 6e-18 |
| $V_{45Gy}$ | 0.08 ± 0.04 | 0.05 ± 0.04 | 1e-20 | 0.09 ± 0.04 | 0.04 ± 0.03 | 3e-19 | 0.27 ± 0.10 | 0.16 ± 0.09 | 1e-15 | 0.14 ± 0.07 | 0.08 ± 0.04 | 6e-18 |
| $V_{50Gy}$ | 0.01 ± 0.02 | 0.01 ± 0.02 | 1e-1 | 0.01 ± 0.02 | 0.01 ± 0.02 | 9e-3 | 0.05 ± 0.07 | 0.05 ± 0.09 | 5e-2 | 0.03 ± 0.04 | 0.02 ± 0.04 | 2e-2 |
| $D_{1\%}$ | 49.37 ± 2.13 | 47.92 ± 3.55 | 4e-5 | 49.55 ± 1.79 | 48.03 ± 6.00 | 6e-6 | 50.13 ± 1.97 | 49.51 ± 5.59 | 1e-2 | 49.89 ± 1.92 | 48.87 ± 5.01 | 1e-3 |
| $D_{2\%}$ | 48.61 ± 2.11 | 46.97 ± 3.77 | 5e-6 | 48.87 ± 1.79 | 46.87 ± 5.53 | 4e-7 | 49.79 ± 1.97 | 48.92 ± 5.15 | 4e-3 | 49.39 ± 1.90 | 48.21 ± 4.62 | 4e-4 |
| Mean (Gy) | 23.75 ± 3.38 | 15.87 ± 3.35 | 2e-21 | 28.23 ± 2.78 | 16.90 ± 3.90 | 2e-21 | 36.68 ± 2.99 | 27.54 ± 5.10 | 4e-21 | 28.96 ± 2.85 | 19.51 ± 3.68 | 2e-21 |

**Table F2**. Dosimetric features at different targeting OARs between VMAT and 4π (P-Val represents P value)

| | Rectum | | | Bladder | | | Small Intestine | | | Bowel Bag | | |
|---|---|---|---|---|---|---|---|---|---|---|---|---|
| | VMAT | 4π | P-Val | VMAT | 4π | P-Val | VMAT | 4π | P-Val | VMAT | 4π | P-Val |
| $V_{5Gy}$ | 0.98 ± 0.06 | 0.96 ± 0.06 | 2e-06 | 1.00 ± 0.00 | 0.88 ± 0.12 | 8e-18 | 0.64 ± 0.17 | 0.67 ± 0.14 | 2e-02 | 0.78 ± 0.18 | 0.78 ± 0.15 | 5e-01 |
| $V_{10Gy}$ | 0.95 ± 0.08 | 0.88 ± 0.12 | 7e-12 | 0.98 ± 0.04 | 0.71 ± 0.18 | 4e-31 | 0.57 ± 0.16 | 0.46 ± 0.16 | 5e-15 | 0.72 ± 0.19 | 0.59 ± 0.18 | 1e-17 |
| $V_{20Gy}$ | 0.86 ± 0.12 | 0.74 ± 0.18 | 2e-09 | 0.79 ± 0.15 | 0.52 ± 0.18 | 7e-28 | 0.40 ± 0.15 | 0.21 ± 0.13 | 9e-19 | 0.54 ± 0.18 | 0.31 ± 0.16 | 9e-19 |
| $V_{30Gy}$ | 0.67 ± 0.17 | 0.62 ± 0.19 | 3e-03 | 0.55 ± 0.16 | 0.42 ± 0.17 | 2e-19 | 0.20 ± 0.11 | 0.12 ± 0.09 | 1e-17 | 0.31 ± 0.14 | 0.21 ± 0.12 | 6e-18 |
| $V_{35Gy}$ | 0.57 ± 0.17 | 0.55 ± 0.19 | 4e-01 | 0.46 ± 0.16 | 0.38 ± 0.16 | 3e-11 | 0.14 ± 0.10 | 0.10 ± 0.08 | 1e-15 | 0.23 ± 0.12 | 0.18 ± 0.11 | 8e-17 |
| $V_{40Gy}$ | 0.45 ± 0.17 | 0.48 ± 0.18 | 2e-02 | 0.38 ± 0.15 | 0.34 ± 0.15 | 2e-05 | 0.10 ± 0.08 | 0.08 ± 0.08 | 9e-12 | 0.18 ± 0.11 | 0.16 ± 0.10 | 7e-12 |
| $V_{45Gy}$ | 0.29 ± 0.16 | 0.35 ± 0.19 | 5e-05 | 0.28 ± 0.14 | 0.26 ± 0.15 | 3e-02 | 0.07 ± 0.07 | 0.05 ± 0.06 | 6e-10 | 0.13 ± 0.10 | 0.11 ± 0.08 | 3e-09 |

| $V_{50Gy}$ | 0.08 ± 0.14 | 0.16 ± 0.18 | 1e-07 | 0.11 ± 0.16 | 0.11 ± 0.13 | 6e-01 | 0.02 ± 0.05 | 0.02 ± 0.05 | 9e-02 | 0.05 ± 0.08 | 0.04 ± 0.07 | 8e-02 |
|---|---|---|---|---|---|---|---|---|---|---|---|---|
| $D_{1\%}$ | 49.85 ± 3.68 | 51.94 ± 2.87 | 2e-06 | 50.85 ± 2.05 | 51.58 ± 4.03 | 3e-01 | 49.27 ± 2.61 | 49.82 ± 8.08 | 1e-01 | 50.50 ± 1.98 | 51.86 ± 7.81 | 3e-01 |
| $D_{2\%}$ | 49.51 ± 3.74 | 51.41 ± 2.91 | 1e-05 | 50.53 ± 2.03 | 51.07 ± 3.76 | 6e-01 | 48.00 ± 3.43 | 47.23 ± 7.91 | 2e-05 | 50.13 ± 2.04 | 50.96 ± 6.42 | 8e-01 |
| Mean (Gy) | 35.21 ± 5.44 | 33.65 ± 7.00 | 7e-03 | 32.92 ± 5.17 | 25.89 ± 7.31 | 2e-26 | 16.70 ± 5.44 | 13.02 ± 5.00 | 5e-17 | 22.32 ± 6.53 | 17.64 ± 6.18 | 1e-17 |